\documentclass[a4paper,10pt]{article}
\usepackage[dvips]{graphicx}
\usepackage{fancyhdr}
\usepackage{latexsym}
\usepackage{amsfonts}
\usepackage{amssymb}
\usepackage{amsthm}
\usepackage{amsmath}
\usepackage{color}
\usepackage{colordvi}
\usepackage{axodraw}
\usepackage{array}
\usepackage{hyperref}
\def\N{N_C}

\def\t{\tau}
\def\n{\nu}

\def\p{\pi}
\def\et{\eta}
\def\CPT{$\chi$PT}
\newcommand{\be}{\begin{equation}}
\newcommand{\ee}{\end{equation}}
\newcommand{\ba}{\begin{array}}
\newcommand{\bea}{\begin{eqnarray}}
\newcommand{\eea}{\end{eqnarray}}

\newcommand{\ket}{\,\rangle}
\newcommand{\bra}{\langle \,}
\begin{document}
\begin{titlepage}
\begin{flushright}
{
UAB-FT-716}
\end{flushright}
\vskip 1.5cm

\begin{center}
{\LARGE \bf Resonance Chiral Lagrangian analysis of $\t^- \to \et^{(\prime)} \p^- \p^0 \n_\t$ decays}
\\[50pt]
{\sc  D.~G\'omez Dumm$^{1}$,
P.~Roig$^{2}$}

\vspace{1.4cm} ${}^1$ IFLP, CONICET $-$ Dpto. de F\'{\i}sica,
Universidad Nacional de La Plata,  \\ C.C.\ 67, 1900 La Plata, Argentina. \\[15pt]
${}^2$ Grup de F\'{\i}sica Te\`orica, Institut de F\'{\i}sica d'Altes Energies,
Universitat Aut\`onoma de Barcelona, E-08193 Bellaterra, Barcelona, Spain.\\[10pt]
\end{center}

\vfill

\begin{abstract}
The hadronization structure of $\t^- \to \et \p^- \p^0 \n_\t$ decays is
analyzed using Chiral Perturbation Theory with resonances, considering only
the contribution of the lightest meson resonances at leading order in the
$1/N_C$ expansion. After imposing the asymptotic behavior of vector spectral
functions ruled by QCD, unknown effective couplings are determined by
fitting the $\t^- \to \et \p^- \p^0 \n_\t$ branching ratio and decay
spectrum to recent data. Predictions for the partner decay $\t^- \to
\eta^\prime \p^- \p^0 \n_\t$ and the low-energy behavior of the cross
section $\sigma (e^+e^-\to\eta\pi^+\pi^-)$ are also discussed.
\end{abstract}
\vspace*{1.0cm}

PACS~: 11.15.Pg, 12.38.-t, 12.39.Fe \\
\hspace*{0.45cm}
Keywords~: Hadron tau decays, chiral Lagrangians, QCD, $1/N$ expansion.
\vfill

\end{titlepage}

\section{Introduction}\label{Intro}

Tau decays represent an ideal benchmark where to analyze diverse topics in
elementary particle physics \cite{RecentTalksandReviews}. In particular,
semileptonic decay channels $\tau^-\to H^- \nu_\tau$, where $H$ is some
hadronic state, allow a rather clean theoretical analysis of the
hadronization of the $V-A$ currents in presence of QCD interactions, since
there is no hadron pollution to the leptonic current. Thus, these processes
provide a suitable tool to find out intrinsic properties of the involved
hadron resonances~\cite{EarlyPapers, Pich:1987qq, Kuhn:1990ad,
Decker:1992kj, Kuhn:1992nz}. In this work we concentrate on the analysis of
$\tau^-\to\eta^{(\prime)} \pi^-\pi^0\nu_\tau$ decays. For these channels the
contributions of scalar and pseudoscalar resonances are expected to be
negligible, since they turn out to be forbidden at tree level by symmetry
arguments, such as $G$-parity conservation. In the limit of isospin symmetry
the corresponding amplitudes are driven by the vector current, allowing a
precise study of the couplings in the odd-intrinsic parity sector.

Concerning the theoretical description, it is well known that in the very
low-energy domain ($E\ll M_{\rho}$, where $M_{\rho}$ is the mass of the
$\rho$(770) meson) Chiral Perturbation Theory ($\chi$PT)~\cite{ChPT} is the
adequate tool to describe hadronic $\tau$ decays~\cite{Colangelo:1996hs}.
However, this approach fails when the invariant mass of the hadronic state
becomes comparable with the mass of the lightest vector and/or axial-vector
resonances, therefore a new strategy is needed in order to enlarge the
domain of applicability of \CPT\ to higher energies. One way out in this
sense is to abandon the Lagrangian approach: one can model $\tau$ decay
amplitudes by taking the lowest order (LO) $\chi$PT results to fix the
normalization of the form factors at low energies, incorporating the
dominant vector and axial--vector meson resonance exchanges by modulating
the amplitudes with {\em ad hoc} Breit--Wigner functions~\cite{EarlyPapers,
Pich:1987qq, Kuhn:1990ad, Decker:1992kj, models}. However, it can be seen
that in the low-energy limit this approach is in general not consistent with
next-to-leading order (NLO) $\chi$PT~\cite{ChPT}, hence the usage of this
procedure to reproduce QCD-ruled amplitudes is
questionable~\cite{Portoles:2000sr, GomezDumm:2003ku}. An alternative
approach is to include the lightest resonances as active degrees of freedom
in the theory. This can be done by adding resonance fields to the $\chi$PT
Lagrangian, without any dynamical assumption~\cite{Ecker:1988te,
Donoghue:1988ed, Ecker:1989yg, Cirigliano:2006hb}. The inclusion of these
fields can be carried out together with an expansion in the inverse of the
number of colors ($\N$)~\cite{LargeNc,Manohar:1998xv, Pich:2002xy,
Cirigliano:2003yq}: at the lowest order in the 1/$\N$ expansion, one gets
from QCD an effective theory that includes a spectrum of infinite zero-width
states. However, we know from phenomenology that resonance widths are
relevant, and that the underlying dynamics is dominated by the lightest
resonances. Hence we consider here a model in which resonance widths are
incorporated, taking into account ---in a way consistent with QCD symmetry
requirements--- only the lightest resonant states that dominate the
processes under study\footnote{The idea of considering a minimal number of
hadronic states that, for a given Green function, satisfy QCD short and long
distance constraints within the large $\N$ limit, has been also considered 
in the context of the so-called minimal hadronic approximation to large-$\N$
QCD~\cite{MHA}.}.

A basic assumption of our approach is that the lightest resonant states are
the dominant ones in low-energy phenomenology. In this way, for a given
process it should be sufficient to introduce only the lightest resonance
multiplet carrying the appropriate quantum numbers, while the inclusion of
higher states can be carried out as a correction~\cite{Kpi,Dumm:2009va}. On
the other hand, the Lagrangian is built upon some fundamental QCD-based
features: the effective interactions have to satisfy QCD symmetries, the
low-energy behavior has to be consistent with \CPT, and the asymptotic
behavior of Green functions and associated form factors has to satisfy QCD
constraints. These requirements imply several relations among the effective
couplings that render the theory predictive. The aim of this work is to
study within this framework the decays $\tau^-\to\eta \pi^-\pi^0\nu_\tau$,
$\tau^-\to\eta^\prime\pi^-\pi^0\nu_\tau$, and the low-energy limit of the
cross section $\sigma(e^+e^-\to\et\pi^+\pi^-)$.

The article is organized as follows: In Sect.~\ref{Theo} we recall how the
$\chi$PT Lagrangian with resonances is built (see
e.g.~Ref.~\cite{Portoles:2010yt}). The relevant hadronic form factors for
the decays under study are given in Sect.~\ref{FFetapipi}. In
Sect.~\ref{Shortdistance} we derive the QCD-ruled high-energy constraints on
the couplings, which reduce the number of unknowns to only four. In
Sect.~\ref{Pheno} we show that two of these unknowns can be bounded from
other phenomenological studies performed within the same framework. In this
way we end up with two unknown couplings, which appear to be highly
correlated~\cite{Roig:2010jp, Chen:2012vw}. The possible values of these
couplings are analyzed by fitting experimental data on the differential
decay distribution of $\tau^-\to\eta \pi^-\pi^0\nu_\tau$ and taking into
account the present upper limit on the branching ratio for
$\tau^-\to\eta^\prime\pi^-\pi^0\nu_\tau$. The low-energy behavior of the
cross section $\sigma(e^+e^-\to\et\pi^+\pi^-)$ is also discussed. Our
conclusions are presented in Sect.~\ref{Conclusions}. Finally, in Appendices
A and B we analyze other possible contributions to the decay amplitudes and
quote some useful isospin relations.

\section{Theoretical framework}\label{Theo}

Our effective Lagrangian is basically ruled by the approximate chiral
symmetry of light-flavored QCD ---which drives the interaction of light
pseudoscalar mesons--- and the SU(3)$_{\rm V}$ assignments of resonance
multiplets~\cite{Ecker:1988te, Ecker:1989yg}. As we will see, for the
processes under consideration it is possible to achieve a good agreement
with present experimental data without the inclusion of excited multiplets.
Moreover, it is seen that vector meson dominance (VMD) turns out to be a
good approximation~\cite{Ecker:1988te}, since spin-zero resonance
contributions vanish at tree level in the very accurate isospin symmetry
limit (see App.~\ref{SPSinglets}). In the case of $\tau$ decays, owing to
the relatively large $\tau$ mass it occurs that several resonances reach
their on-shell condition when the amplitudes are integrated over the full
phase space. The corresponding pole singularities can be regularized by
including finite (energy-dependent) resonance widths, thus departing from
the lowest order in the $1/N_C$ expansion. Here we adopt the prescription in
Ref.~\cite{GomezDumm:2000fz}, where energy-dependent resonance widths have
been calculated in a well-defined way using our Lagrangian formalism.

We will work out $\t^-\to\et^{(\prime)}\p^-\p^0\n_\t$ decays considering
exact isospin symmetry. In this limit the processes are driven only by the
vector current (see Sect.~\ref{FFetapipi}), and appear to be dominated by
the contributions of the $\rho(770)$ resonance. The relevant effective
Lagrangian reads~:
\begin{eqnarray}
\label{eq:ret} {\cal L}_{\rm R\chi T}   & \doteq   & {\cal L}_{WZW} \,+ \,
{\cal L}_{\rm kin}^{\rm V}\, + \, \frac{F^2}{4}\langle u_{\mu} u^{\mu} + \chi _+
\rangle \, + \, \frac{F_V}{2\sqrt{2}} \langle V_{\mu\nu} f_+^{\mu\nu}\rangle
\, \nonumber \\
& &  \hspace{-1.9cm} + \ i \,\frac{G_V}{\sqrt{2}} \langle V_{\mu\nu} u^\mu
u^\nu\rangle \, +\, \sum_{i=1}^{7}  \, \frac{c_i}{M_V}  \, {\cal O}^i_{\rm
VJP} \,+\, \sum_{i=1}^{4}  \, d_i  \, {\cal O}^i_{\rm VVP} \,+\,
\sum_{i=1}^{5}  \, \frac{g_i}{M_V}  \, {\cal O}^i_{\rm VPPP} \ ,
\label{lagrangian}
\end{eqnarray}
where all coupling constants are real, $F$ and $M_V$ being the pion decay
constant and the mass of the lightest vector meson resonances, respectively.
We follow here the notation in
Refs.~\cite{GomezDumm:2003ku,Ecker:1988te,RuizFemenia:2003hm}\footnote{In
Ref.~\cite{Chen:2012vw}, two additional operators (${\cal \tilde{O}}^8_{\rm
VJP}$ and ${\cal \tilde{O}}^5_{\rm VVP}$) have been found when the singlet
$\left\langle VVP\right\rangle$ Green function is considered in addition to
the octet one in the $p^2\sim m_q\sim1/N_C$ counting. In
App.~\ref{SPSinglets} we show that they do not contribute to the hadronic
tau decays studied here.}. Accordingly, $\langle \rangle$ stands for trace
in flavor space, and $u^\mu$, $\chi_+$ and $f_+^{\mu\nu}$ are defined by
\begin{eqnarray}
u^\mu & = & i\,u^\dagger\, D^\mu U\, u^\dagger \,,\nonumber \\
\chi_\pm & = & u^\dagger\, \chi \, u^\dagger\, \pm u\,\chi^\dagger\, u\,, \nonumber \\
f_\pm^{\mu\nu} & = & u^\dagger\, F_L^{\mu\nu}\, u^\dagger\, \pm
u\,F_R^{\mu\nu}\, u \ ,
\end{eqnarray}
where $u$ ($U=u^2$), $\chi$ and $F_{L,R}^{\mu\nu}$ are $3\times 3$ matrices
that contain light pseudoscalar fields, current quark masses and external
left and right currents, respectively. The matrix $V^{\mu\nu}$ includes the
lightest vector meson multiplet, and ${\cal L}_{\rm kin}^{\rm V}$ stands for
the resonance kinetic term. The first term in Eq.~(\ref{lagrangian}) is the
Wess-Zumino-Witten interaction Lagrangian~\cite{Wess:1971yu,Witten:1983tw},
which governs the decay amplitudes studied here in the limit of low hadron
momenta. The part of this interaction that contributes to the processes
considered here reads
\begin{eqnarray}
\mathcal{L}_{WZW} & \doteq & -\frac{i N_C}{48 \pi^2}\; \epsilon_{\mu\nu\alpha\beta}
\; \langle \Sigma^\mu_L\, U^\dagger \,\partial^\nu r^\alpha\, U \,l^\beta\;
+ \;\Sigma^\mu_L \, l^\nu \,\partial^\alpha l^\beta \nonumber \\
& & + \;\Sigma^\mu_L\,
\partial^\nu l^\alpha\, l^\beta- (L \leftrightarrow R) \rangle\ ,
\end{eqnarray}
where $\Sigma_{L,R}$ are given by $\Sigma^\mu_L=
U^\dagger\partial^\mu U$, $\Sigma^\mu_R = U\partial^\mu U^\dagger$, and
$l^\alpha$ and $r^\alpha$ are left and right external currents.
 Finally, the operators ${\cal O}_{\rm VJP}^i$, ${\cal
O}_{\rm VVP}^i$ and ${\cal O}_{\rm VPPP}^i$ in Eq.~(\ref{lagrangian}) are
given by~:

\hfill

$VJP$ terms
\begin{eqnarray}
{\cal O}_{\rm VJP}^1 & = & \epsilon_{\mu\nu\rho\sigma}\,
\langle \, \{V^{\mu\nu},f_{+}^{\rho\alpha}\} \nabla_{\alpha}u^{\sigma}\,\rangle
\; \; , \nonumber\\[3mm]
{\cal O}_{\rm VJP}^2 & = & \epsilon_{\mu\nu\rho\sigma}\,
\langle \, \{V^{\mu\alpha},f_{+}^{\rho\sigma}\} \nabla_{\alpha}u^{\nu}\,\rangle
\; \; , \nonumber\\[3mm]
{\cal O}_{\rm VJP}^3 & = & i\,\epsilon_{\mu\nu\rho\sigma}\,
\langle \, \{V^{\mu\nu},f_{+}^{\rho\sigma}\}\, \chi_{-}\,\rangle
\; \; , \nonumber\\[3mm]
{\cal O}_{\rm VJP}^4 & = & i\,\epsilon_{\mu\nu\rho\sigma}\,
\langle \, V^{\mu\nu}\,[\,f_{-}^{\rho\sigma}, \chi_{+}]\,\rangle
\; \; , \nonumber\\[3mm]
{\cal O}_{\rm VJP}^5 & = & \epsilon_{\mu\nu\rho\sigma}\,
\langle \, \{\nabla_{\alpha}V^{\mu\nu},f_{+}^{\rho\alpha}\} u^{\sigma}\,\rangle
\; \; ,\nonumber\\[3mm]
{\cal O}_{\rm VJP}^6 & = & \epsilon_{\mu\nu\rho\sigma}\,
\langle \, \{\nabla_{\alpha}V^{\mu\alpha},f_{+}^{\rho\sigma}\} u^{\nu}\,\rangle
\; \; , \nonumber\\[3mm]
{\cal O}_{\rm VJP}^7 & = & \epsilon_{\mu\nu\rho\sigma}\,
\langle \, \{\nabla^{\sigma}V^{\mu\nu},f_{+}^{\rho\alpha}\} u_{\alpha}\,\rangle
\;\; ;
\label{eq:VJP}
\end{eqnarray}

$VVP$ terms
\begin{eqnarray}
{\cal O}_{\rm VVP}^1 & = & \epsilon_{\mu\nu\rho\sigma}\,
\langle \, \{V^{\mu\nu},V^{\rho\alpha}\} \nabla_{\alpha}u^{\sigma}\,\rangle
\; \; , \nonumber\\[3mm]
{\cal O}_{\rm VVP}^2 & = & i\,\epsilon_{\mu\nu\rho\sigma}\,
\langle \, \{V^{\mu\nu},V^{\rho\sigma}\}\, \chi_{-}\,\rangle
\; \; , \nonumber\\[3mm]
{\cal O}_{\rm VVP}^3 & = & \epsilon_{\mu\nu\rho\sigma}\,
\langle \, \{\nabla_{\alpha}V^{\mu\nu},V^{\rho\alpha}\} u^{\sigma}\,\rangle
\; \; , \nonumber\\[3mm]
{\cal O}_{\rm VVP}^4 & = & \epsilon_{\mu\nu\rho\sigma}\,
\langle \, \{\nabla^{\sigma}V^{\mu\nu},V^{\rho\alpha}\} u_{\alpha}\,\rangle
\; \; ;
\label{eq:VVP}
\end{eqnarray}

$VPPP$ terms
\begin{eqnarray}
{\cal O}_{\rm VPPP}^1 & = & i \, \varepsilon_{\mu\nu\alpha\beta} \,
\left\langle V^{\mu\nu} \, \left( \, h^{\alpha\gamma} u_{\gamma} u^{\beta}
- u^{\beta} u_{\gamma} h^{\alpha\gamma} \right) \right\rangle \, ,
\nonumber  \\ [2mm] {\cal O}_{\rm VPPP}^2 & = & i \,
\varepsilon_{\mu\nu\alpha\beta} \, \left\langle V^{\mu\nu} \, \left( \,
h^{\alpha\gamma} u^{\beta} u_{\gamma} - u_{\gamma} u^{\beta}
h^{\alpha\gamma} \, \right) \right\rangle \, , \nonumber \\ [2mm] {\cal
O}_{\rm VPPP}^3 & = & i \, \varepsilon_{\mu\nu\alpha\beta} \, \left\langle
V^{\mu\nu} \, \left( \, u_{\gamma} h^{\alpha\gamma} u^{\beta}  -
u^{\beta} h^{\alpha\gamma} u_{\gamma} \, \right) \right\rangle \, ,
\nonumber \\ [2mm] {\cal O}_{\rm VPPP}^4 & = &
\varepsilon_{\mu\nu\alpha\beta} \, \left\langle \left\lbrace  \,V^{\mu\nu}
\,,\,  u^{\alpha}\, u^{\beta}\, \right\rbrace \,{\cal \chi}_{-}
\right\rangle \, , \nonumber \\ [2mm] {\cal O}_{\rm VPPP}^5 & = &
\varepsilon_{\mu\nu\alpha\beta} \, \left\langle
 \,u^{\alpha}\,V^{\mu\nu} \, u^{\beta}\, {\cal \chi}_{-} \right\rangle \, ,
\label{eq:VPPP}
\end{eqnarray}
where $h_{\mu\nu} = \nabla_\mu u_\nu + \nabla_\nu u_\mu$. The covariant
derivative $\nabla_\mu$ involves pseudoscalar meson fields and $l^\alpha$,
$r^\alpha$ external currents. Its explicit expression can be found in
Ref.~\cite{Ecker:1988te}.

The nonet of vector resonances $V$ is described here using the antisymmetric
tensor formulation. In the context of VMD~\cite{Ecker:1989yg}, this is shown
to be consistent with the usage of the $\chi$PT Lagrangian for light
pseudoscalar mesons up to ${\cal O}(p^2)$ in the even-intrinsic parity
sector and up to ${\cal O}(p^4)$ in the odd-intrinsic parity
sector~\cite{Cirigliano:2006hb}.

\section{Form factors in $\t^-\to\et^{(\prime)}\p^- \p^0 \n_\t$}\label{FFetapipi}

In the Standard Model, $\tau^- \rightarrow \et^{(\prime)}\pi^- \p^0
\nu_{\tau}$ decay amplitudes can be written as
\begin{equation}
{\cal M}  \, = \,  - \, \frac{G_F}{\sqrt{2}} \, V_{ud} \, \bar u_{\nu_\tau}
\gamma^\mu\, (1-\gamma_5) u_\tau\, \mathcal{H}_\mu \; ,
\end{equation}
where $V_{ud}\simeq \cos\theta_C$ is the relevant Cabibbo-Kobayashi-Maskawa
mixing and $\mathcal{H}_\mu$ is the hadron matrix element of the left-handed
QCD current $(V-A)_{\mu}$. In general, for a decay of a $\tau$ lepton into
three pseudoscalar mesons the hadronic tensor $\mathcal{H}_\mu$ can be
written as~\cite{Kuhn:1992nz}
\begin{eqnarray} \label{generaldecomposition_3mesons}
\!\!\!\!\!\!\!\bra h_1(p_1)h_2(p_2)h_3(p_3)|(V-A)^\mu|0\ket & = &
F_1^A(Q^2,\,s_1,\,s_2)\, V_1^\mu\, \nonumber \\
& & \hspace{-5.4cm} + \; F_2^A(Q^2,\,s_1,\,s_2)\,V_2^\mu\;
+\; i\, F_3^V(Q^2,\,s_1,\,s_2)\,V_3^\mu\; +
\; F_4^A(Q^2,\,s_1,\,s_2)\, Q^\mu \, ,
\end{eqnarray}
where
\begin{eqnarray} \label{Set_of_independent_Vectors_3meson}
& & V_1^\mu\, = \, \left( g^{\mu\nu} - \frac{Q^{\mu}Q^{\nu}}{Q^2}\right) \,
(p_1 - p_3)_{\nu} \,\,\, , \quad V_2^\mu\, = \, \left( g^{\mu\nu} -
\frac{Q^{\mu}Q^{\nu}}{Q^2}\right) \,
(p_2 - p_3)_{\nu} \, ,\nonumber\\
& & V_3^\mu\, =
\,\varepsilon^{\mu\alpha\beta\gamma}\,p_{1\alpha}\,p_{2\beta}\,p_{3\gamma}\,\,\,,\quad
Q^\mu\,=\,(p_1\,+\,p_2\,+\,p_3)^\mu\,,\quad s_i\,=\,(Q\,-\,p_i)^2\ .
\nonumber\\
\end{eqnarray}
The upper indices in the form factors indicate the participating currents,
either the axial-vector ($A$), or the vector one ($V$). The form factors
$F_1^A$ and $F_2^A$ drive a transition to hadronic states with quantum
numbers $J^P\,=\,1^+$, while $F_3^V$ and $F_4^A$ correspond to outgoing
states with $J^P\,=\,1^-$ and $J^P\,=\,0^-$, respectively.
Let us focus on the amplitude for the transition $\tau^{-} \rightarrow
\eta_8(p_1)\, \pi^{-}(p_2) \, \pi^{0}(p_{3})\, \nu_{\tau}$, considering
the limit of exact isospin symmetry. First of all, it is easy to see that
for this process the axial-vector form factors vanish from $G$-parity
conservation, therefore the dynamics will be essentially determined by the
form factor $F_3^V$.

{}From the effective Lagrangian in Eq.~(\ref{lagrangian}), the diagrams that
contribute to $F_3^V$ are those represented in Fig.~1, where single solid
lines correspond to $\pi$ and $\eta$ mesons and double lines to the
$\rho(770)$ resonance.
\begin{figure}
\begin{center}
\includegraphics[scale=0.7]{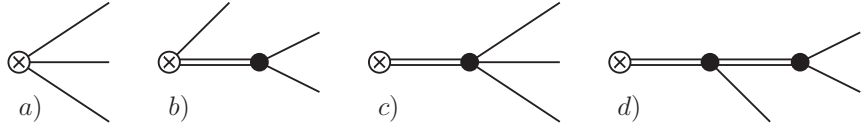}
\caption[]{\label{fig_diagsetapipi} Topologies contributing to the final
hadron state in $\tau^- \rightarrow \et^{(\prime)} \pi^- \pi^0 \,
\nu_{\tau}$ decays in the $N_C \rightarrow \infty$ limit. Crossed circles
indicate QCD vector current insertions. Single lines represent pseudoscalar
mesons ($\et$, $\pi$)  while double lines stand for $\rho$-resonance
intermediate states.}
\end{center}
\end{figure}
The corresponding contributions to the vector form factor read
\begin{eqnarray}
\label{T_etapipi_chi}
{F_3^{V(a)}}_{(\eta_8\pi\pi)} & = &  \frac{N_C }{6\,\sqrt{6} \, \pi^2 \, F^{3}}\,, \\
\label{T_etapipi_1R1}
{F_3^{V(b)}}_{(\eta_8\pi\pi)} & = & \frac{8 \, G_V}{\sqrt{3}\,F^3 \, M_V} \,\frac{1}{M_\rho^2-s_1}
\left[ c_{125}\, Q^2 \right. -\; c_{1256} \, s_1  \nonumber \\
 & & \left. +\; c_{1235}\, m_\eta^2 +
 8 c_3\left(m_\pi^2-m_\eta^2\right) \right]\,, \\
\label{T_etapipi_1R2}
{F_3^{V(c)}}_{(\eta_8\pi\pi)} & = & -\;\frac{16\, F_V}{\sqrt{3}\,M_V \,F^3} \frac{1}{M_\rho^2 - Q^2} \,
\left[ \, g_{123}\, s_1\,\right. -\,g_2\, \left(\, Q^2\,+\,2 m_\pi^2\,-\,m_\eta^2\right) \nonumber \\
& & \left. -\, \left(g_1\,-\,g_3\right)\,2\,m_\pi^2\,
+\,g_{45} \,m_\pi^2 \right]\,, \\
{F_3^{V(d)}}_{(\eta_8\pi\pi)} & = & -\;\frac{8 \sqrt{2}}{\sqrt{3}} \, \frac{F_V G_V}{F^3} \,
\frac{1}{M_\rho^2-Q^2}\;\frac{1}{M_\rho^2-s_1}
\nonumber\\
\label{T_etapipi_2R}
& & \times\; \left[ d_3\,(Q^2+s_1) + \, (d_{12}-d_3)\,
m_\eta^2\, + \, 8d_2\, (m_\pi^2 - m_\eta^2) \right]\, ,
\end{eqnarray}
where we have defined
\begin{eqnarray}
c_{125} & = & c_1-c_2+c_5\,, \nonumber \\
c_{1256} & = & c_1-c_2-c_5+2c_6\,, \nonumber \\
c_{1235} & = & c_1+c_2+8c_3-c_5\,, \nonumber \\
g_{123} & = & g_1+2g_2-g_3\,, \nonumber \\
g_{45} & = & 2g_4+g_5\,, \nonumber \\
d_{12} & = & d_1+8d_2\,.
\end{eqnarray}

The amplitude for the $\tau$ decay into the $\eta_0\pi^-\pi^0$ hadronic
state can be read from Eqs.~(\ref{T_etapipi_chi}) to (\ref{T_etapipi_2R}) by
simply multiplying ${F_3^{V(a,b,c,d)}}_{(\eta_8\pi\pi)}$ by $\sqrt{2}$.
Then, the matrix elements for the decays into the physical hadronic states
$\eta\pi^-\pi^0$ and $\eta^\prime\pi^-\pi^0$ can be obtained by considering
$\eta_8-\eta_0$ mixing. Here we will consider a double angle mixing
scheme~\cite{DoubleAngleMixing}, which is consistent with the large-$N_C$
expansion~\cite{Kaiser:1998ds}. Using a notation similar to that in
Ref.~\cite{Escribano:2010wt}, the SU(3) octet and singlet fields are
collected in a doublet $\eta_B^T\equiv (\eta_8,\eta_0)$, while the physical
fields are included in $\eta_P^T\equiv (\eta,\eta^\prime)$. These doublets
are related by the transformation $\eta_B=(\mathcal{M})^T\eta_P$,
where~\cite{Escribano:2010wt}
\begin{equation}
\label{NLOmixing}
\mathcal{M}=\left(
\begin{array}{lr}
\cos\theta_P(1-\delta_8/2)+\sin\theta_P\,\delta_{80}/2 &\
-\sin\theta_P(1-\delta_0/2)-\cos\theta_P\,\delta_{80}/2 \\[1ex]
\sin\theta_P(1-\delta_8/2)-\cos\theta_P\,\delta_{80}/2 &\
\cos\theta_P(1-\delta_0/2)-\sin\theta_P\,\delta_{80}/2
\end{array}
\right)\ .
\end{equation}
In the framework of R$\chi$T, the parameters $\delta_8$, $\delta_0$ and
$\delta_{80}$ can be in fact derived from an effective Lagrangian that
involves scalar resonances~\cite{Ecker:1988te}. If the latter are
organized in a U(3) matrix $S$, from the lowest order Lagrangian
\begin{equation}
\mathcal{L}^S=c_d\langle S u_\mu u^\mu\rangle + c_m\langle
S\chi_+\rangle\, \label{1SLagrangian}
\end{equation}
one gets
\begin{eqnarray}
\delta_8=\frac{8 c_d c_m}{M_S^2}\frac{M_8^2}{F^2}\ ,\qquad
\delta_0=\frac{8 c_d c_m}{M_S^2}\frac{M_0^2}{F^2}\ ,\qquad \delta_{80}
=\frac{8 c_d c_m}{M_S^2}\frac{M_{80}^2}{F^2}\, ,
\end{eqnarray}
where\footnote{The fully dominant contribution to the $\eta^\prime$ mass is
not due to current quark masses but to the $U(1)_A$
anomaly~\cite{etaprimemass}, through the topological susceptibility of
gluondynamics. Hence we keep $\theta_P$ as a free parameter, to be fitted
from phenomenology.}
\begin{eqnarray}
M_8^2 & = & \frac{1}{3}\left(4M_K^2-M_\pi^2\right)\ ,\nonumber\\
M_0^2
 & = & \frac{1}{3}\left(2M_K^2+M_\pi^2\right)\ ,\nonumber \\
M_{80}^2 & = & -\frac{2\sqrt{2}}{3}\left(M_K^2-M_\pi^2\right)\, .
\end{eqnarray}
Here we take for $M_\pi$ and $M_K$ the isospin averaged values of the pion
and kaon masses, neglecting higher order corrections in the combined
chiral and $1/N_C$ expansion. In addition we assume $c_d c_m = F^2/4$
\cite{Jamin:2000wn, Jamin:2001zq}, which is required by high-energy QCD in the
$N_C\to\infty$ limit. Finally, from $\eta-\eta'$ phenomenology we take
$M_S\simeq 0.980$ GeV and
$\theta_P=(-13.3\pm0.5)^\circ$~\cite{Ambrosino:2009sc}.

Given the form factors, $F_3^V(Q^2,s_1,s_2)$, the spectral functions for
the decays $\t^-\to\et^{(\prime)}\p^-\p^0\n_\t$ are finally given by
\begin{eqnarray}
\label{Q2 spectrum}
 \frac{{\rm d}\Gamma}{{\rm d}Q^2} & = &
 \frac{G_F^2|V_{ud}|^2}{128(2\pi)^5M_\tau}\left(\frac{M_\tau^2}{Q^2}-1\right)^2
 \frac{1}{3}\left(1+2\frac{Q^2}{M_\tau^2}\right) \nonumber \\
 & & \times \int_{(m_\eta+m_\pi)^2}^{(\sqrt{Q^2}-m_\pi)^2}{\rm d}s_2
 \int_{t_-(Q^2,s_2)}^{t_+(Q^2,s_2)}{\rm d}s_1\ W_B(Q^2,s_1,s_2)\,,
\end{eqnarray}
where the relevant structure function $W_B$~\cite{Kuhn:1992nz} is defined by
$W_B(Q^2,s_1,s_2)=V_3^2\,|F_3^V(Q^2,s_1,s_2)|^2$ and the limits of the integral
over $s_1$ are
\begin{eqnarray}
 t_{\pm}(Q^2,s_2) & = &
 \frac{1}{4s_2}\ \bigg\{\left(Q^2+m_\eta^2-2m_\pi^2\right)^2 \nonumber \\
& &
- \left[\lambda^{1/2}(Q^2,s_2,m_\pi^2)\mp\lambda^{1/2}(m_\eta^2,m_\pi^2,s_2)
\right]^2 \bigg\}\ ,
\end{eqnarray}
with $\lambda(a,b,c)=(a+b-c)^2-4ab$. We have neglected here the neutrino mass.

\section{Short distance constraints on the couplings}\label{Shortdistance}

The above form factors depend on several combinations of coupling constants,
besides the $\rho$ mass and the pion decay constant. The values of these
parameters are not provided by the effective theory, and their determination
from the underlying QCD theory is still an open problem. However, one can
get information on the effective couplings by assuming that the resonance
region provides a bridge between the chiral and perturbative regimes, even
when one does not include the full resonance spectrum~\cite{Ecker:1989yg}.
This is implemented by matching the high-energy behavior of Green functions
(or related form factors) evaluated within the resonance Lagrangian with
asymptotic results obtained in perturbative QCD~\cite{GomezDumm:2003ku,
Ecker:1989yg, Cirigliano:2006hb, MHA, Dumm:2009va, Chen:2012vw,
RuizFemenia:2003hm, QCDResonances}. In particular, it has been shown that
the analysis of the two-point Green functions
$\Pi_{A,\,V}$~\cite{Ecker:1989yg} and the three-point Green function $VVP$
of QCD currents (with the inclusion of only one multiplet of vector
resonances)~\cite{RuizFemenia:2003hm} leads to the following constraints in
the $\N\to\infty$ limit:
\begin{itemize}
\item[i)] By demanding that the two-pion vector form factor vanishes at
high momentum transfer one obtains the condition $F_V \, G_V =
F^2$~\cite{Ecker:1989yg}.
\item[ii)] The analysis of the $VVP$ Green function~\cite{RuizFemenia:2003hm}
leads to the following results for the couplings in Eqs.
(\ref{T_etapipi_1R1}), (\ref{T_etapipi_1R2}) and (\ref{T_etapipi_2R})~: \bea
\label{eq:condVVP}
c_{125} &=& 0\,, \nonumber\\
c_{1235} &=& 0\,, \nonumber\\
c_{1256} &=& -\frac{N_C}{32\pi^2}\frac{M_V}{\sqrt{2}F_V}\,, \nonumber\\
d_{12} &=& -\frac{N_C}{64\pi^2} \frac{M_V^2}{F_V^2} \, + \, \frac{F^2}{4 F_V^2}\,, \nonumber\\
d_3 &=& -\frac{N_C}{64\pi^2}\frac{M_V^2}{F_V^2} \, + \, \frac{F^2}{8F_V^2}\ .
\eea
\end{itemize}

On the other hand, it is possible to find additional constraints by
requiring that the contributions of any intermediate hadronic state to the
spectral function Im$\Pi_V(Q^2)$ vanish in the limit $Q^2\to \infty$. This
is a reasonable assumption, since from perturbative QCD Im$\Pi_V(Q^2)$ has
to go to a constant value for $Q^2\to\infty$ \cite{Floratos:1978jb}, and the
imaginary part of the two-point Green function can be understood as the sum
of infinite intermediate hadronic states. Considering the
intermediate $\eta^{(\prime)}\pi\pi$ hadronic states one gets the following
constraints on the coupling constants:
\begin{eqnarray}\label{rels}
c_{125} &  = & 0\,, \nonumber \\
c_{1256} & = &
- \, \frac{N_C}{96 \pi^2} \, \frac{M_V\,F_V}{\sqrt{2} \, F^2}\,, \nonumber \\
d_3 & = & - \, \frac{N_C}{192 \pi^2} \frac{M_V^2}{F^2} \nonumber\,,  \\
g_{123} & = & 0 \,,\nonumber \\
g_2\, & = & \, \frac{N_C}{192 \pi^2}\, \frac{M_V}{\sqrt{2} \,F_V} \; .
\end{eqnarray}
It is worth to notice that relations~(\ref{rels}) are in agreement with
those found in a similar analysis carried out for $\tau$ decays into $2K
\pi\nu_\tau$~\cite{Dumm:2009kj, Roig:2007yp} and $P^-\gamma\nu_\tau$
($P=\pi,K$) states~\cite{GuoRoig}. Comparing with Eqs.~(\ref{eq:condVVP}),
we agree in the vanishing of $c_{125}$, while the constraints for $c_{1256}$
and $d_3$ cannot be simultaneously satisfied keeping agreement with their
values in Eq.~(\ref{rels}). Moreover, as stated in Ref.~\cite{Dumm:2009kj},
it is seen that the expected vanishing of the $\pi\gamma^\star\gamma$ form
factor at high-$q^2$ is obtained from Eqs.~(\ref{rels}) but not from
Eqs.~(\ref{eq:condVVP}). In any case, numerically the differences are small,
and the impact of these couplings on the observables is rather
mild\footnote{The introduction of additional resonances has a different
effect on the short-distance relations obtained from the $VVP$ Green
function and from the imaginary part of the vector--vector correlator. While
all new contributions to the correlator are positive definite, this is not
true for the $VVP$ Green function, where cancellations are allowed. Thus the
outcome of both procedures may be different when the spectrum is restricted
to the lowest--lying resonances. A convergence of both results should be
recovered if the full tower of excited resonances is taken into account.}.
Thus we choose to stick to our set of relations (\ref{rels}), using
Eqs.~(\ref{eq:condVVP}) to fix the combinations $c_{1235}$ and $d_{12}$, not
obtained within our study. In this way, the analysis of short distance
constraints allows to reduce significantly the number of unknown coupling
constants in the form factors $F_3^{V(\alpha)}$ quoted in
Eqs.~(\ref{T_etapipi_chi}-\ref{T_etapipi_2R}). To calculate the decay
amplitudes for the processes $\tau^-\to\eta^{(\prime)}\pi^-\pi^0\nu_\tau$ we
end up with just four unknown parameters, namely $F_V$, $c_3$, $g_{45}$ and
$d_2$. As in the above mentioned analysis, we will take $M_V= M_\rho$.

\section{Phenomenological analysis}\label{Pheno}

In order to carry out a phenomenological analysis of
$\tau^-\to\eta^{(\prime)}\pi^-\pi^0\nu_\tau$ decays we take into account the
available experimental information. In the case of
$\tau^-\to\eta\pi^-\pi^0\nu_\tau$ this includes the measured branching
fraction BR$(\tau^-\to\eta\pi^-\pi^0\nu_\tau) = (1.39\pm 0.10) \times
10^{-3}$~\cite{PDG}, as well as the data on the corresponding spectral
function obtained by Belle~\cite{Inami:2008ar}. The process $\tau^{-}
\rightarrow \eta^\prime \pi^{-} \pi^{0} \nu_{\tau}$ has not been observed
yet, hence we consider only the upper bound given by the PDG~\cite{PDG},
namely BR$(\tau^{-} \rightarrow \eta^\prime \pi^{-} \pi^{0} \nu_{\tau})<
8.0\times 10^{-5}$ at 90$\%$ confidence level.

As stated, the number of unknown parameters entering the vector form factor
$F_3^V$ in R$\chi$T can be reduced to four by means of the short-distance
constraints obtained in Sect.~\ref{Shortdistance}. In addition, the values
of $F_V$ and $g_{45}$ can be estimated within R$\chi$T from the
phenomenological analysis of $\t^-\to (\p\p\p)^-\n_\t$ and
$\omega\to\pi^+\pi^-\pi^0$, respectively: the best fit to the $\t^-\to
(\p\p\p)^-\n_\t$ spectral function measured by ALEPH~\cite{Barate:1998uf}
corresponds to $F_V=0.180$ GeV~\cite{Dumm:2009va} with an estimated error of
$\sim15\%$\footnote{Some theoretical analyses lead to the value
$F_V=\sqrt{3}F\sim
0.160$~GeV~\cite{GuoRoig,Guo:2007hm,Pich:2010sm,Nieves:2011gb,Guo:2011pa,Guo:2012ym,Guo:2012yt}.
We have checked that a change of $F_V$ within the range $[0.160,0.180]$~GeV
does not affect significantly the results presented in this section.}, while
from the $\omega\to\pi^+\pi^-\pi^0$ branching ratio one gets $g_{45} =
-0.60\pm 0.02$~\cite{Dumm:2009kj}. In this way we are left with only two
unknowns, namely the coupling constants $c_3$ and $d_2$. Our goal is to be
able to describe the available experimental information just by fitting
these two parameters.

\begin{figure}[h!]
\begin{center}
\vspace*{1.2cm}
\includegraphics[scale=0.9]{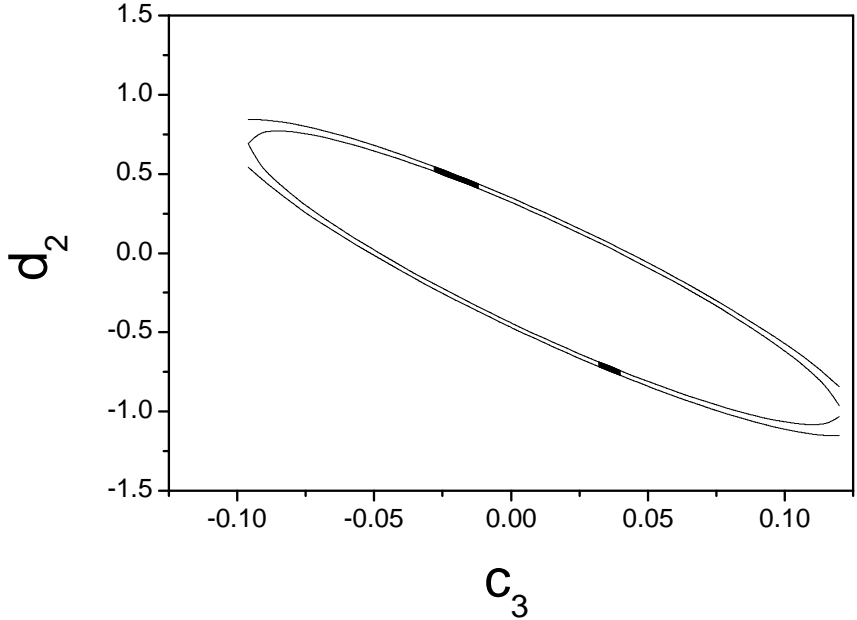}
\caption[]{\label{Fig:d2c3} \small{Contour in the $c_3-d_2$ plane
compatible with the branching ratio of the decay $\tau^{-} \rightarrow
\eta \pi^{-} \pi^{0} \nu_{\tau}$ at the level of one standard deviation.
The black strips highlight the two regions that yield lowest values of
$\chi^2/dof$ (within one unit) from a fit to Belle
data~\cite{Inami:2008ar} on the corresponding spectral function.}}
\end{center}
\end{figure}

\begin{figure}[h!]
\begin{center}
\vspace*{0.9cm}
\includegraphics[scale=0.5,angle=-90]{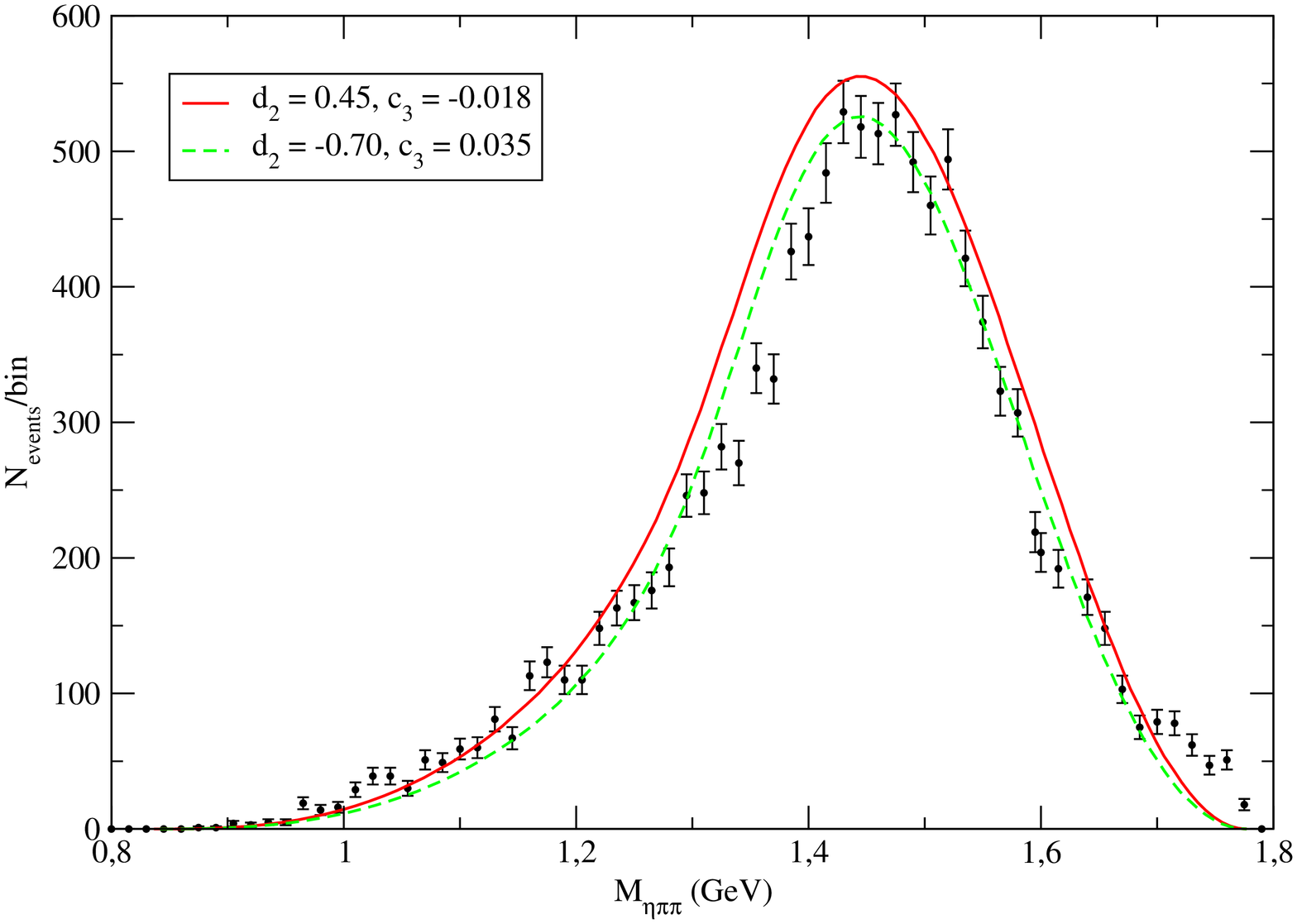}
\caption[]{\label{Fig:spectrumetapipi} \small{Theoretical curves fitting
the spectral function for $\tau^{-} \rightarrow \eta \pi^{-} \pi^{0}
\nu_{\tau}$ decay, compared to experimental data~\cite{Inami:2008ar}.}}
\end{center}
\end{figure}

In Figure \ref{Fig:d2c3} we show the $c_3-d_2$ parameter region compatible
with the PDG branching ratio for the mode $\tau^{-} \rightarrow \eta \pi^{-}
\pi^0 \nu_{\tau}$ at the level of one sigma. A large correlation between
both couplings can be appreciated, in agreement with
Refs.~\cite{Roig:2010jp, Chen:2012vw}\footnote{In particular, as noticed in
Ref.~\cite{Chen:2012vw}, there is an anticorrelation between $c_3-d_2$ in
$\tau^{-}\rightarrow \eta^\prime \pi^{-} \pi^0 \nu_{\tau}$ and in associated
radiative decays. Therefore, the combined study could improve the
determination of these couplings.}. Then, taking into account this allowed
region for $c_3$ and $d_2$, we have carried out a fit to Belle
data~\cite{Inami:2008ar} for the $\tau^-\to\eta\pi^-\pi^0\nu_\tau$ spectral
function. We find two $\chi^2$ minima, located at $(c_3,d_2) =
(-0.018,0.45)$ and $(c_3,d_2) = (0.035,-0.70)$, with $\chi^2/{\rm dof} =
3.80$ and $4.34$, respectively, where the statistical error is about $10\%$.
These values are indicated in Fig.~\ref{Fig:d2c3}, where black strips
correspond to the $c_3-d_2$ regions that keep $\chi^2/{\rm dof}$ within one
unit far from the minima. The corresponding theoretical curves for the
spectral function, together with experimental data, are shown in
Fig.~\ref{Fig:spectrumetapipi}. We have also carried out a fit to the
normalized spectral function, obtaining that the preferred values for
$(c_3,d_2)$ remain almost unchanged, while $\chi^2/{\rm dof}$ values get
reduced to 3.0 for both minima. Finally, in order to account for the
theoretical error of the high-energy predictions for the couplings
$c_{1256}$, $d_3$, $g_2$ and $d_{12}$, we have also fitted the data allowing
these coupling combinations to vary within $\pm 1/3$ of the values obtained
from Eqs.~(\ref{eq:condVVP}) and (\ref{rels})\footnote{We have also
considered nonvanishing values for the coupling $c_{1235}$, which should be
zero according to Eq.~(\ref{eq:condVVP}). Notwithstanding, have kept
$c_{125}$ and $g_{123}$ equal to zero. Indeed, if $c_{125}\neq 0$ the
Brodsky-Lepage behavior~\cite{Brodsky-Lepage} of the form factor is
violated, and Im$\Pi_V$ goes, asymptotically, as $Q^6$ $\log(Q^2/M_V^2)$; if
$g_{123}\neq 0$, the asymptotic growth goes as $\mathcal{O}(Q^6)$. Varying
$c_{1235}$ in the range $\left[-0.05,0.05\right]$ does not improve the
fit.}. Noteworthy, the $\chi^2$ value does not get reduced, which can be
taken as an indication that our short-distance relations (obtained at
leading order in $1/N_C$) lead to an appropriate effective Lagrangian to
reproduce the experimental observations.

\begin{figure}[h!]
\begin{center}
\vspace*{0.9cm}
\includegraphics[scale=0.8]{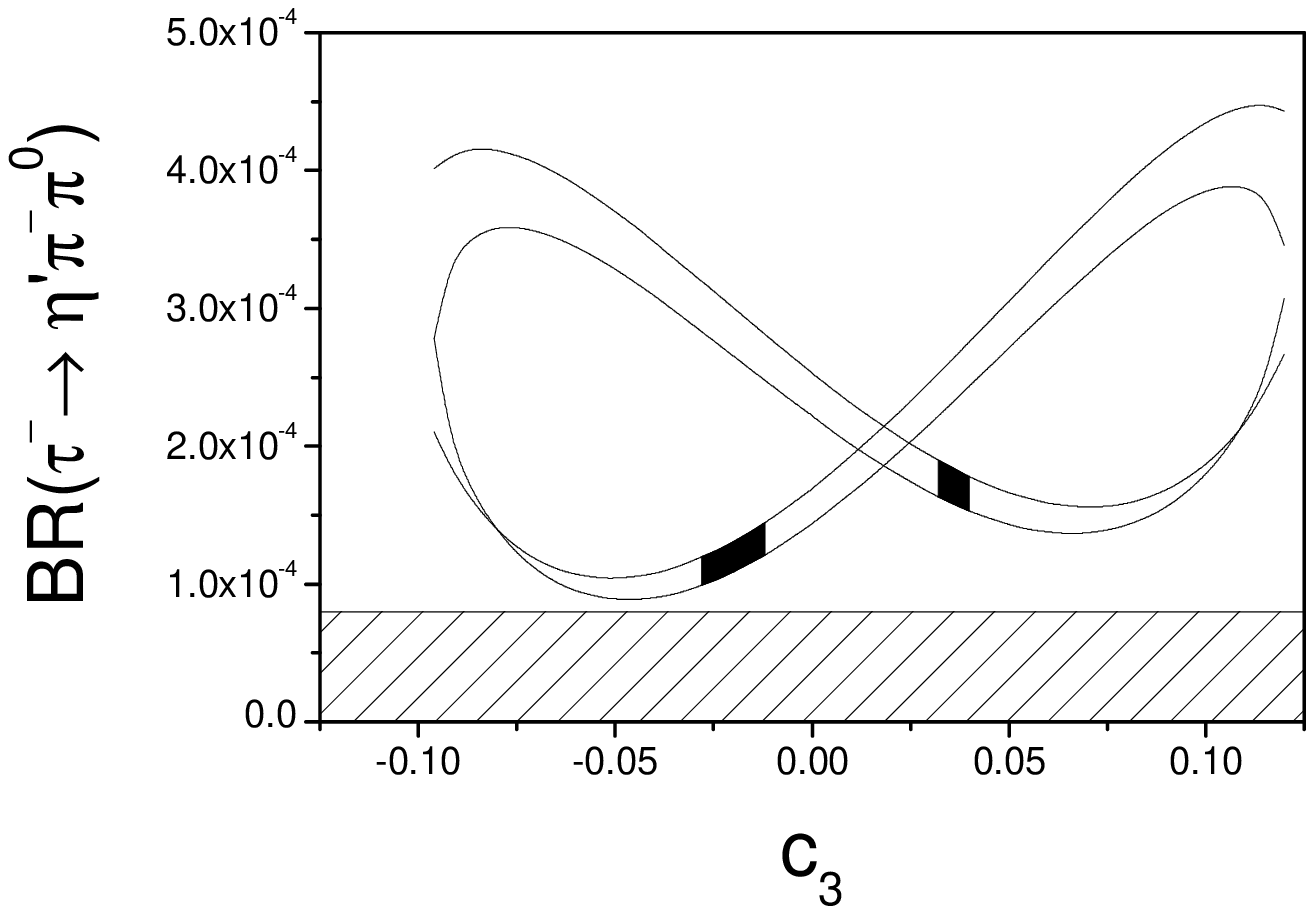}
\caption[]{\label{Fig:bretaprima} \small{Prediction for the branching ratio
of the decay $\tau^{-} \rightarrow \eta^\prime \pi^{-} \pi^{0} \nu_{\tau}$
consistent (within one sigma) with the $\tau^{-} \rightarrow \eta \pi^{-}
\pi^{0} \nu_{\tau}$ branching ratio quoted by the PDG. The horizontal line,
corresponding to BR$(\tau^{-} \rightarrow \eta^\prime \pi^{-} \pi^{0}
\nu_{\tau}) = 0.8\times 10^{-4}$, represents the current PDG bound. The
notation for the black strip is the same as in Fig.~\ref{Fig:d2c3}.}}
\end{center}
\end{figure}

Considering the fitted values for $c_3$ and $d_2$, we can calculate the
corresponding predictions for the $\tau^{-} \rightarrow \eta^\prime
\pi^{-} \pi^0 \nu_{\tau}$ branching ratio. The results are shown in
Fig.~\ref{Fig:bretaprima}, where we have taken $c_3$ as the independent
parameter. It is seen that the predictions are somewhat above the 90\%
confidence level upper bound quoted by the PDG, which is indicated by the
shadowed region in the figure. However, the result corresponding to $c_3 =
-0.018$ turns out to be rather close to the upper bound; in fact,
compatibility is achieved if the width of the $c_3-d_2$ band is enlarged
considering two standard deviations in the measured value of BR$(\tau^{-}
\rightarrow \eta \pi^{-} \pi^0 \nu_{\tau})$. Future, more precise
measurements of the $\tau^{-} \rightarrow \eta^\prime \pi^{-} \pi^0
\nu_{\tau}$ process should indicate whether our slight discrepancy arises
from a weakness in the theoretical assumptions (e.g.\ treatment of
$\eta_8-\eta_0$ mixing, effect of excited resonances, SU(3) breaking terms
in the Lagrangian~\cite{Moussallam:2007qc})
or it just reflects an issue in the detection of this $\tau$ decay mode. In this
regard, we emphasize the importance of making global fits with unified and
consistent treatments of all hadronic currents, in order to avoid
cross-contamination between different hadronic tau decay channels from
misunderstood backgrounds. The improvement in the most relevant hadronic
matrix elements in TAUOLA~\cite{TAUOLA,Olga} may be a key tool in this
sense.

\begin{figure}[h!]
 \begin{center}
  \vspace*{1.1cm}
\includegraphics[scale=0.5, angle=-90]{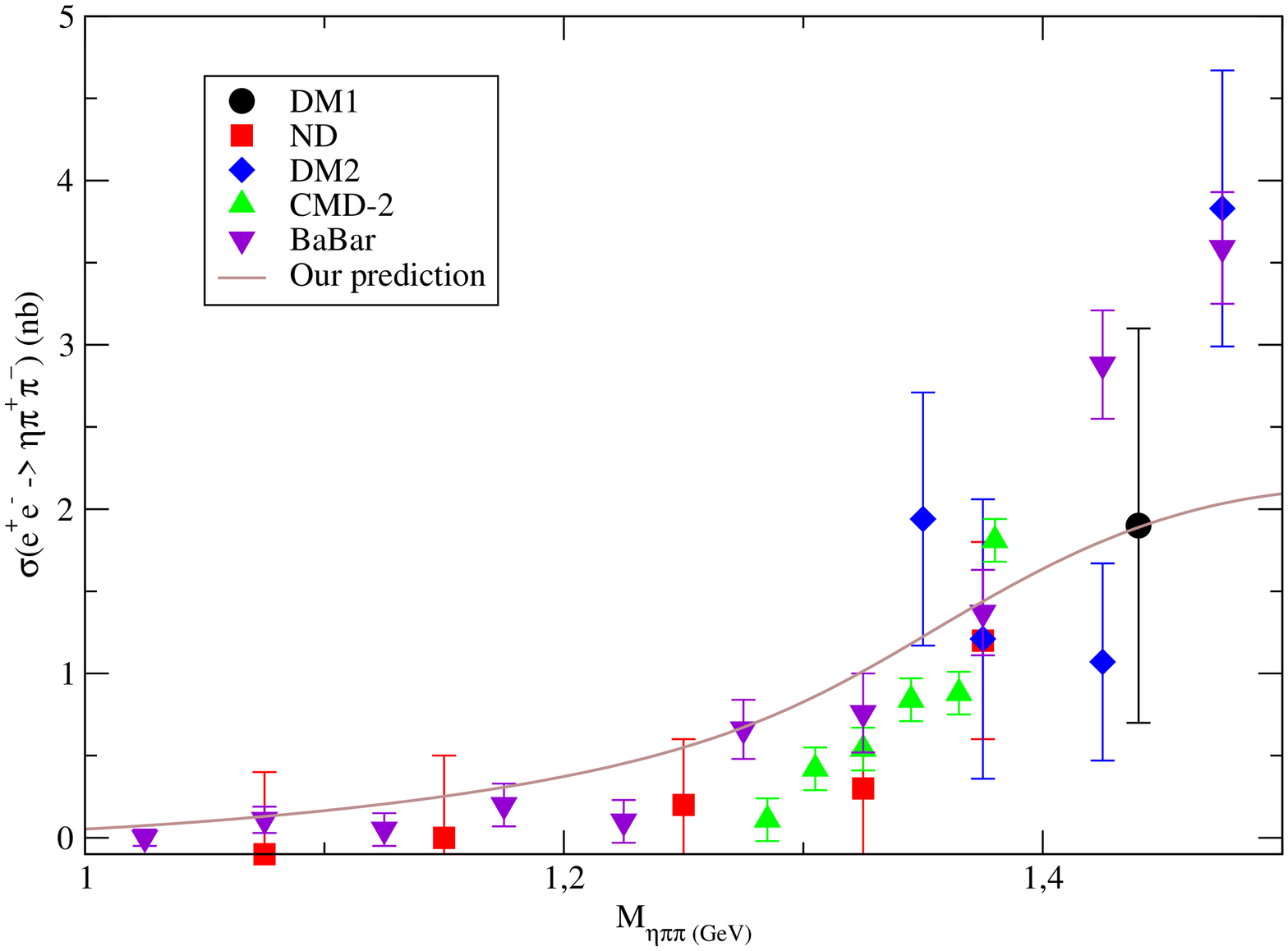}
\caption[]{\label{Fig:Comparisontoe+e-data} \small{Prediction for
$\sigma(e^+e^-\to \eta\pi^+\pi^-)$ low-energy behavior from our analysis of
$\tau^{-} \rightarrow \eta \pi^{-} \pi^{0} \nu_{\tau}$ decays, in comparison
with DM1 \cite{Delcourt:1982sj}, ND \cite{Druzhinin:1986dk}, DM2
\cite{Antonelli:1988fw}, CMD-2 \cite{Akhmetshin:2000wv} and BaBar
\cite{Aubert:2007ef} data.}}
 \end{center}
\end{figure}

Finally, our analysis can be used to predict
$\sigma(e^+e^-\to\eta\pi^+\pi^-)$ in the low energy region (conversely,
one could in general use data on $e^+e^-$ annihilation into hadronic
states to get predictions for the corresponding semileptonic tau decays
\cite{Eidelman:1990pb, Cherepanov:2009zz}\footnote{A more elaborated
dedicated approach, also based in R$\chi$T, has been developed for
$\sigma(e^+e^-\to\eta/\pi^0\, \pi^+\pi^-)$~\cite{OlgaJorge}.}). The
relation between this cross section and the $\tau^{-} \rightarrow \eta
\pi^{-} \pi^{0} \nu_{\tau}$ spectral function is detailed in Appendix
\ref{isos}. One gets
\begin{equation} \label{Eq.Iso}
 \frac{d\Gamma(\tau^-\to \eta\pi^-\pi^0\nu_\tau)}{\mathrm{d}Q^2}  =
  2\, f(Q^2) \sigma( e^+e^-\to \eta\pi^+\pi^-)\,,
\end{equation}
where $f(Q^2)$ is given by
\begin{equation} \label{fQ2}
 f(Q^2)=\frac{G_F^2 |V_{ud}|^2}{384(2\pi)^5
M_\tau}\left(\frac{M_\tau^2}{Q^2}-1\right)^2\left(1+2\frac{Q^2}{M_\tau^2}\right)
\left(\frac{\alpha^2}{48\pi}\right)^{-1}Q^6\, .
\end{equation}
In Fig.~\ref{Fig:Comparisontoe+e-data} we quote our predictions for the
cross section, in comparison with low-energy $e^+e^-$ data obtained in
various experiments~\footnote{We note that in this neutral current process
there are additional contributions from new operators ${\cal O}^8_{\rm
VJP}$ and ${\cal O}^5_{\rm VVP}$, see Ref.~\cite{Chen:2012vw}, which
implies the introduction of two additional unknown couplings. However,
these terms are suppressed in the large-$N_C$ limit in the standard
counting (see discussion in App.~\ref{SPSinglets}).}. We notice that
although the $\eta^\prime$ meson decays to $\eta \pi^+ \pi^-$ with a
fraction of about $45\%$, there is no significant contamination from the
chain $\sigma(e^+e^-\to\eta^\prime\gamma^\star\to\eta\pi^+\pi^-)$ since,
due to $C$ parity, this occurs at NLO in powers of the electromagnetic
coupling $\alpha$. From the figure it is seen that our results are
consistent with experimental data up to a center of mass energy of about
1.4 GeV. In fact, one should not expect our treatment to be valid beyond
this energy region, where effects of excited states should be sizeable and
there is no phase space suppression as in $\tau$ decay spectral functions.

\section{Conclusions}
\label{Conclusions} We have worked out the decays $\tau^{-} \rightarrow \eta
\pi^{-} \pi^0 \nu_{\tau}$ and $\tau^{-} \rightarrow \eta^\prime \pi^{-}
\pi^0 \nu_{\tau}$ within the framework of  of Chiral Perturbation Theory
with resonances. The theoretical analysis has been based on the large-$N_C$
expansion of QCD, the low-energy limit given by \CPT\ and the appropriate
asymptotic behavior of the form factors, which helps to fix most of the
initially unknown effective couplings. Indeed, after taking into account
information acquired in the previous related studies, $\tau^{-} \rightarrow
\eta^{(\prime)} \pi^{-} \pi^0 \nu_{\tau}$ amplitudes can be written in terms
of only two unknown parameters.

We have carried out a phenomenological analysis taking into account the
experimental data for the branching ratio and the spectrum of the decay
$\tau^{-} \rightarrow \eta \pi^{-} \pi^0 \nu_{\tau}$, as well as the
present upper bound for the $\tau^{-} \rightarrow \eta^\prime \pi^{-}
\pi^0 \nu_{\tau}$ branching fraction. A fit to the data allows to
determine two preferred sets of values for the unknown parameters $c_3$
and $d_2$ in the effective Lagrangian, namely $(c_3,d_2) = (-0.018,0.45)$
and $(0.035,-0.88)$, which lead to a reasonable overall description of the
spectrum. The former set seems to be favored by the predictions for the
branching ratio BR$(\tau^{-} \rightarrow \eta^\prime \pi^{-} \pi^0
\nu_{\tau})$, although in both cases the theoretical values appear to be
somewhat above the present experimental upper bound. Finally, using
isospin symmetry, these results can be used to get a prediction for the
low-energy behavior of the cross section $\sigma(e^+e^-\to
\eta\pi^+\pi^-)$. The results are in good agreement with the available
experimental information, and the approach can be useful \cite{OlgaJorge}
for the implementation of the related hadronic current in the
PHOKHARA~\cite{Rodrigo:2001kf} Monte Carlo generator.

Our present results should be regarded as a first step in the study of the
$\tau^{-} \rightarrow \eta^{(\prime)} \pi^{-} \pi^0 \nu_{\tau}$ decays in
our framework. In the light of higher statistics for the
$\et\pi^-\pi^0\nu_\tau$ mode, or the observation of the decay $\tau^{-}
\rightarrow \eta^{\prime} \pi^{-} \pi^0 \nu_{\tau}$, this description
could be improved by considering e.g.\ the exchange of excited vector
resonances, SU(3) breaking terms in the Lagrangian
or revising the $\eta_8-\eta_0$ mixing scheme.

\section*{Acknowledgments}
We are indebted to Antonio Pich for his critical reading of our draft. We
acknowledge Jorge Portol\'es for making useful remarks on this project. We
appreciate very much that K.~Hayasaka and K.~Inami gave us access to Belle
data for our research. We wish to thank Henryk Czyz and Simon Eidelman since
their interest in the low-energy description of
$\sigma\left(e^+e^-\to\et\pi^+\pi^-\right)$ triggered, in fact, this part of
our work. We also acknowledge Simon's help in accessing the experimental
data and Henryk's comments putting forward a relevant consequence of our
isospin analysis. It is always encouraging and rewarding for P.~R.~ to
collaborate with T.~Przedzinski, O.~Shekhovtsova and Z.~Was. P.~R.~ has
benefited from discussions with R.~Escribano and P.~Masjuan on
$\eta-\eta^\prime$ mixing and with V.~Cherepanov and I.~M.~Nugent on the
decay modes and cross-section studied in this article. This work has been
partially supported by the Spanish grants FPA2007-60323, FPA2011-25948 and
by the Spanish Consolider Ingenio 2010 Programme CPAN (CSD2007-00042). It
has also been founded in part by CONICET and ANPCyT (Argentina), under
grants PIP02495 and PICT07-03-00818.

\appendix
\renewcommand{\theequation}{\Alph{section}.\arabic{equation}}
\renewcommand{\thetable}{\Alph{section}.\arabic{table}}
\setcounter{section}{0}
\setcounter{equation}{0}
\setcounter{table}{0}

\section{Contribution of spin-zero resonances and singlet terms}
\label{SPSinglets}

In this Appendix we analyze both the contribution of spin-zero (scalar and
pseudoscalar) resonances and SU(3) singlet couplings to $\tau^{-}
\to\eta^{(\prime)} \pi^{-} \pi^{0} \nu_{\tau}$ decays.

\subsection{Scalar and pseudoscalar resonance exchange}

Discrete symmetries of QCD constrain the possible couplings in the effective
Lagrangian. One of these symmetries is G-parity, which is exact in the SU(2)
symmetry limit. With our conventions, the corresponding quantum numbers are:
$G_{A_\mu}=-1$, $G_{V_\mu}=+1$, $G_\eta=+1$, $G_{\pi^{(\star)}}=-1$,
$G_{f_0/\sigma}=+1$, $G_{a_0}=-1$, thus the final state $\eta\pi^-\pi^0$ has $G =
+$. As stated in Sect.~3, since the axial-vector weak current has $G=-$, only
the vector current can contribute to the
$\t^-\to\et^{(\prime)}\p^-\p^0\n_\t$ decay amplitudes in this limit. The
intermediate states $f_0\pi^-$, $\sigma\pi^-$ are also forbidden by G-parity
conservation, which only leaves the channels $a_0^-\pi^0\to\eta\pi^-\pi^0$
and $a_0^0\pi^-\to\eta\pi^0\pi^-$. However, the vector current is $J^P=1^-$,
while the intermediate states $a_0\pi$ have parity $P = (-1)^{J+1}$.
Therefore, one can conclude that both scalar and pseudoscalar resonance
contributions are strongly suppressed at tree level and can be safely neglected.

\subsection{Contribution of double-trace terms}

In Ref.~\cite{Chen:2012vw} two additional operators have been found with
respect to those in Ref.~\cite{RuizFemenia:2003hm}. Although these operators
involve two traces, hence they are suppressed in the standard counting (in
powers of $p^2\sim m_q^2$ and $1/N_C$), it is seen that they become leading
when a simultaneous counting in all three expansion parameters is carried
out~\cite{Kaiser:1998ds}. The operators read
\begin{eqnarray}
 \tilde{O}^8_{VJP} & = & -i \tilde{c}_8 M_V \,\sqrt{\frac{2}{3}}\,
 \varepsilon_{\mu\nu\rho\sigma}\left\langle V^{\mu\nu}\tilde{f}_+^{\rho\sigma}\right\rangle
 \log (\mathrm{det}\,\tilde{u}) \nonumber\\
 \tilde{O}^5_{VVP} & = & -i \tilde{d}_5 M_V^2 \,\sqrt{\frac{2}{3}}\,
 \varepsilon_{\mu\nu\rho\sigma}\left\langle V^{\mu\nu} V^{\rho\sigma}\right\rangle
 \log (\mathrm{det}\,\tilde{u})\ ,
\end{eqnarray}
where the tildes stand for $u$ and $f$ matrices that include the singlet
term (and would contribute to the processes considered here through the
$\eta$ and $\eta^\prime$ components of the $\eta_0$ meson). Once again the
contribution of these operators vanishes, since the second operator only
contributes to neutral current processes, while the first one leads to the
contraction of symmetric and antisymmetric tensors in the $\tau$ decay
amplitudes.

\appendix
\renewcommand{\theequation}{\Alph{section}.\arabic{equation}}
\renewcommand{\thetable}{\Alph{section}.\arabic{table}}
\setcounter{section}{1}
\setcounter{equation}{0}
\setcounter{table}{0}

\section{Isospin relations}
\label{isos} In this Appendix we provide a derivation of Eq.~(\ref{Eq.Iso}),
which allows to relate the $\tau^{-} \to\eta \pi^{-} \pi^{0} \nu_{\tau}$
differential decay rate and the $\sigma(e^+e^-\to \eta\pi^+\pi^-)$ cross
section.

We work in the limit of SU(2) isospin symmetry, and we neglect $Z$-exchange
contributions to the hadronic $e^+e^-$ cross-section, which is a safe
approximation in the considered energy range. Thus this process will be
driven by the vector current, via photon exchange. One expects to get a
relation between this cross section and the vector current contribution to
the decay of a tau lepton into the corresponding hadronic state.

Since both $\eta_8$ and $\eta_0$ states are SU(2) singlets, we can compute
isospin relations between $\eta_{0,8}\pi\pi$ channels just by taking into
account the isospin of $\pi\pi$ states. Let us denote by $T_{-0}$, $T_{0-}$
the amplitudes $\bra \eta\pi\pi|\overline{d}\Gamma^\mu_w u|0\ket$ and by
$T_{+-},T_{-+},T_{00}$ the amplitudes $\frac{1}{\sqrt 2}\langle
\eta\pi\pi|(\bar u\Gamma^\mu_w u-\bar d\Gamma^\mu_w d)|0 \rangle$, where the
subscripts correspond to pion electric charges, and $\eta$ can be either
$\eta_0$ or $\eta_8$. We obtain the relations
\begin{eqnarray}
& & \frac{1}{\sqrt{2}}\left(T_{-0}+T_{0-}\right) = -\frac{1}{\sqrt{6}}\left(
T_{+-}+T_{-+}-2T_{00}\right) = 0\ ,
\nonumber\\
& & \frac{1}{\sqrt{2}}\left( T_{0-}-T_{-0}\right) =
-\frac{1}{\sqrt{2}}\left( T_{-+}-T_{+-}\right)\ ,
\nonumber\\
& &  \sqrt{3}\left( T_{+-}+T_{-+}-T_{00}\right) = 0 \ ,
\end{eqnarray}
which lead to
\begin{equation}
 T_{00}=0\;\;,T_{+-}=-T_{-+}=T_{0-}=-T_{-0}\ .
\end{equation}
Now let us consider the electromagnetic current. One can
decompose it into $I=0$ and $I=1$ pieces:
\begin{equation}
\Gamma^\mu_{em} = \frac{1}{3}\left( 2\overline{u}\gamma^\mu
u-\overline{d}\gamma^\mu d-\overline{s}\gamma^\mu s\right) =
\Gamma^\mu_{(0)}+\Gamma^\mu_{(1)}\,,
\end{equation}
where
\begin{equation}
\Gamma^\mu_{(0)} = \frac{1}{6}\left(\overline{u}\gamma^\mu
u+\overline{d}\gamma^\mu d-2\overline{s}\gamma^\mu
s\right)\,,\,\,\Gamma^\mu_{(1)}=\frac{1}{2}\left(\overline{u}\gamma^\mu
u-\overline{d}\gamma^\mu d\right)\, .
\end{equation}
One can relate the amplitudes $\bra \eta\pi\pi|\Gamma^\mu|0\ket$ for charge
and isospin ($A_I$) $|\eta\pi\pi\ket$ states by
\begin{equation}
A_{+-}=\frac{1}{\sqrt{2}}A_1+\frac{1}{\sqrt3}A_0\,,\quad
A_{-+}=-\frac{1}{\sqrt{2}}A_1+\frac{1}{\sqrt3}A_0\,,\quad
A_{00}=-\frac{1}{\sqrt{3}}A_0\,.
\end{equation}
Moreover the vanishing of the amplitude $A_2$ implies
\begin{equation}
 2A_{00}+A_{+-}+A_{-+}=0\,.
\end{equation}
In this way one obtains the following relations:
\begin{eqnarray}
A_{+-}+A_{00}=\frac{1}{\sqrt{2}}A_1\,,\;
A_{-+}+A_{00}=-\frac{1}{\sqrt{2}}A_1\,,\;
A_1=\frac{A_{+-}-A_{-+}}{\sqrt{2}}\,,
\nonumber\\
A_0=\frac{A_{+-}+A_{-+}-A_{00}}{\sqrt{3}}=-\sqrt{3}A_{00}=\frac{\sqrt{3}}{2}\left(
A_{+-}+A_{-+}\right) \,,
\label{b7}
\end{eqnarray}
which lead to
\begin{equation}
|A_{+-}+A_{-+}|^2+|A_{+-}-A_{-+}|^2= 2\left(
|A_{+-}|^2+|A_{-+}|^2\right)=4|A_{00}|^2+2|A_1|^2\,,
\end{equation}
\begin{equation}
 |A_1|^2=|A_{+-}|^2+|A_{-+}|^2-2|A_{00}|^2\,.
\end{equation}
Thus corresponding cross sections are related by
\begin{eqnarray} \label{I=1 component}
\!\!\!\!\!\!\!\sigma\left( e^+e^-\to \eta\pi\pi\right)|_{I=1} & =&
\sigma\left( e^+e^-\to \eta\pi^+\pi^-\right)+\sigma\left( e^+e^-\to \eta\pi^-\pi^+\right) \nonumber \\
 & & -\ 2\times2\;\sigma\left( e^+e^-\to
\eta\pi^0\pi^0\right)\,\nonumber \\
& \simeq & 2\,\sigma\left( e^+e^-\to \eta\pi^+\pi^-\right) \ ,
\end{eqnarray}
where the additional factor $2$ in the first equation arises from the
presence of identical particles in the final state. In the last line we have
neglected the cross section to the $\eta\pi^0\pi^0$ state, since it turns
out to vanish at the lowest order in the electromagnetic coupling $\alpha$
owing to $C$-parity conservation. For the isoscalar part, from Eq.~(\ref{b7}) we find
\begin{equation}\label{I=0 component}
 \sigma\left( e^+e^-\to \eta\pi\pi\right)|_{I=0} = 6\, \sigma\left(
e^+e^-\to \eta\pi^0\pi^0\right)\ .
\end{equation}
Finally, one has
\begin{equation}
 \frac{1}{\sqrt{2}}\left(T^{0-}-T^{-0}\right)=\sqrt{2}\, T^{0-}=-\langle
1,0|\frac{\bar{u}u-\bar{d}d}{\sqrt{2}} |0\rangle =-\sqrt{2}\, A_1\,.
\end{equation}
Taking into account that the $e^+e^-$ cross-section into three hadrons is given by
\begin{equation} \label{e+e-xsect3hads}
\sigma_{e^+e^-\to h_1h_2h_3}(Q^2)=\frac{e^4}{768\,\pi^3}\frac{1}{Q^6}\int
\mathrm{d}s \;\mathrm{d}t\, |F_3|^2 \left(-V_{3\mu} V_3^{\mu *}\right)\,,
\end{equation}
the cross-sections for the different modes read ($|A_{+-}|^2 = |A_{1}|^2/2 =
|A_{-0}|^2/2$)
\begin{eqnarray}
 \sigma\left( e^+e^-\to \eta\pi^+\pi^-\right) & = &
\frac{\alpha^2}{96\pi}\frac{1}{Q^6} \int \mathrm{d}s\; \mathrm{d}t\;
|T_{-0}|^2 \left( V_{3\mu} V^{3\mu *}\right) \,,
 \nonumber\\
 \sigma\left( e^+e^-\to \eta\pi^0\pi^0\right) & =  &
\frac{\alpha^2}{48\pi}\frac{1}{Q^6} \int \mathrm{d}s\; \mathrm{d}t\;
\frac{1}{2}|T_{00}|^2 \left( V_{3\mu} V^{3\mu *}\right)\ ,
\end{eqnarray}
where the additional factor $1/2$ in the second equation arises from the
presence of identical particles in the final state. Thus one finally obtains
\begin{eqnarray}
 \frac{d\Gamma(\tau^-\to \eta\pi^-\pi^0\nu_\tau)}{\mathrm{d}Q^2} & = &
 f(Q^2) \, \sigma(e^+e^-\to \eta\pi\pi)|_{I=1} \nonumber \\
 & = & 2 f(Q^2) \left[\sigma( e^+e^-\to \eta\pi^+\pi^-) - 2\,\sigma
 (e^+e^-\to \eta\pi^0\pi^0)\right] \nonumber \\
 & \simeq & 2 f(Q^2) \,\sigma( e^+e^-\to \eta\pi^+\pi^-)\ ,
\end{eqnarray}
where $f(Q^2)$ is the kinematical factor given in Eq.~(\ref{fQ2}).

\end{document}